\documentclass[12pt]{article}
\usepackage{times}
\usepackage{geometry}
\usepackage[utf8]{inputenc}
\usepackage{enumitem,amssymb}
\usepackage{ragged2e}
\usepackage{graphicx}
\newlist{thematic}{itemize}{8}
\setlist[thematic]{label=$\square$}
\usepackage{pifont}

\begin{document}
\raggedright
\huge
Astro2020 Science White Paper \linebreak

Stellar physics with high-resolution UV spectropolarimetry\linebreak
\normalsize

\noindent \textbf{Thematic Areas:} \hspace*{60pt} $\square$ Planetary Systems \hspace*{10pt} $\blacksquare$ Star and Planet Formation \hspace*{20pt}\linebreak
$\square$ Formation and Evolution of Compact Objects \hspace*{31pt} $\square$ Cosmology and Fundamental Physics \linebreak
  $\blacksquare$  Stars and Stellar Evolution \hspace*{1pt} $\blacksquare$ Resolved Stellar Populations and their Environments \hspace*{40pt} \linebreak
  $\square$    Galaxy Evolution   \hspace*{45pt} $\square$             Multi-Messenger Astronomy and Astrophysics \hspace*{65pt} \linebreak
  
\textbf{Principal Author:} 

Name: Coralie Neiner
 \linebreak						
Institution: LESIA, Paris Observatory, CNRS, PSL University, Sorbonne Universit\'e, Univ. Paris Diderot, Sorbonne Paris Cit\'e, 5 place Jules Janssen, 92195 Meudon, France
 \linebreak
Email: coralie.neiner@obspm.fr
 \linebreak
Phone:  +33145077785
 \linebreak
 
\textbf{Co-authors:} 

Julien Morin (Laboratoire Univers et Particules de Montpellier (LUPM), Universit\'e de Montpellier, CNRS, 34095 Montpellier, France),\\
Jean-Claude Bouret (Aix Marseille Univ, CNRS, CNES, LAM, Marseille, France),\\
Luca Fossati (Space Research Institute, OeAW, Graz, Austria)
  \linebreak

\justify
\textbf{Abstract:}
Current burning issues in stellar physics, for both hot and cool stars, concern their magnetism. In hot stars, stable magnetic fields of fossil origin impact their stellar structure and circumstellar environment, with a likely major role in stellar evolution. However, this role is complex and thus poorly understood as of today. It needs to be quantified with high-resolution UV spectropolarimetric measurements. In cool stars, UV spectropolarimetry would provide access to the structure and magnetic field of the very dynamic upper stellar atmosphere, providing key data for new progress to be made on the role of magnetic fields in heating the upper atmospheres, launching stellar winds, and more generally in the interaction of cool stars with their environment (circumstellar disk, planets) along their whole evolution.

\pagebreak
\section{Massive and hot stars}

\indent Massive stars provide heavy chemical elements to the Universe and dominate the interstellar radiation field. Moreover, they are the progenitors of  core-collapse supernovae and gamma-ray bursts, leaving behind compact objects such as neutron stars and black holes. These, when in binary systems, may trigger the emission of gravitational waves during coalescence. In addition, due to their luminosity and spectroscopic features, the successive phases of massive stars and starbursts can be observed out to large distances. Therefore, they are essential for many domains of astrophysics, such as stellar and planetary formation and galactic structure and evolution. It is therefore crucial to understand the physical processes at work in massive stars.\\
\indent Many advances have already occurred in the past regarding, e.g., the role of rotation in the structure and evolution of massive stars. More recently, thanks to high precision photometric missions such as TESS, asteroseismology has also provided new insights into the interior of those stars. However, one physical ingredient remains very poorly understood in massive stars and hot stars in general: their magnetic field, and its impact on their structure, evolution, and environment.\\
\indent About 10\% of O, B, and A stars host a magnetic field of fossil origin, usually dipolar but inclined with respect to the stellar rotation axis, with a polar field strength ranging from a few hundreds to a few ten thousands Gauss (Neiner et al. 2015; Grunhut \& Neiner 2015). The $\sim$90\% of stars that do not host such a field may nevertheless host an ultra-weak field of the order of 1 Gauss, such as those recently discovered in some A and Am stars (e.g. Blazère et al. 2016). The presence of a magnetic field, even a weak one, is critical for stellar structure and evolution as it is expected to have a strong impact on the circumstellar environment.\\
\indent The current burning issues about massive stars therefore concern their magnetism and are the following:
\begin{itemize}
    \item Why are there only $\sim$10\% of magnetic hot stars with a field above a few hundreds of Gauss? Since the magnetic fields are of fossil origin, i.e. descendants from a seed field present in the molecular cloud from which the star was formed and enhanced by a dynamo during the early phases of the life of the star (when it was fully convective), the occurrence rate of magnetic fields is likely related to the initial conditions of stellar formation. Quantitatively relating the initial conditions to the presence of a stable field would give us important insight into the early phases of stellar evolution.
    \item Why are there less magnetic hot stars in short-period binary systems ($\sim$2\%; Alecian et al. 2015) than among single stars ($\sim$10\%)? This is also probably related to the fossil origin of magnetic fields in hot stars, possibly to the difficulty to fragment cores in a magnetized medium (Commercon et al. 2011). Understanding this occurrence rate difference between single and binary hot stars would cast new light on star formation.
    \item How do fossil magnetic fields evolve and how do they impact the evolution of the star during and after the main sequence? Is magnetic flux conserved throughout the life of the star or are there some processes at work producing a decay (e.g. Fossati et al. 2016) or an enhancement of the field during stellar evolution?
    \item Do dynamo fields develop in the convective regions that appear in the radiative envelope in the second part of the stellar life? If so, what is the result of the interaction between the fossil field and the dynamo field(s) on stellar structure and evolution (see e.g. Featherstone et al. 2009)?
    \item How do magnetic fields interact with fluid motions inside the star, impact the internal rotation profile, the transport of chemical elements and of angular momentum, and thus impact stellar evolution?
    \item How do magnetic fields impact the mass-loss rate of hot stars, in particular through the confinement of the wind particles into magnetospheres around the stars and magnetic braking (e.g. Meynet et al. 2011)? In addition, mass loss quenching through magnetic fields provides an explanation to form heavy stellar mass Black Holes with masses $>$ 25 M$_\odot$ such as those inferred from the detection of gravitational waves (Petit et al., 2017).
    \item How do magnetic fields contribute to the energy budget and type of supernovae explosions (Guilet et al. 2010) and modify the very end stages of stellar evolution (e.g. formation of magnetars)?
\end{itemize}

\begin{figure*}[t]
    \centering
    \includegraphics[height=8cm]{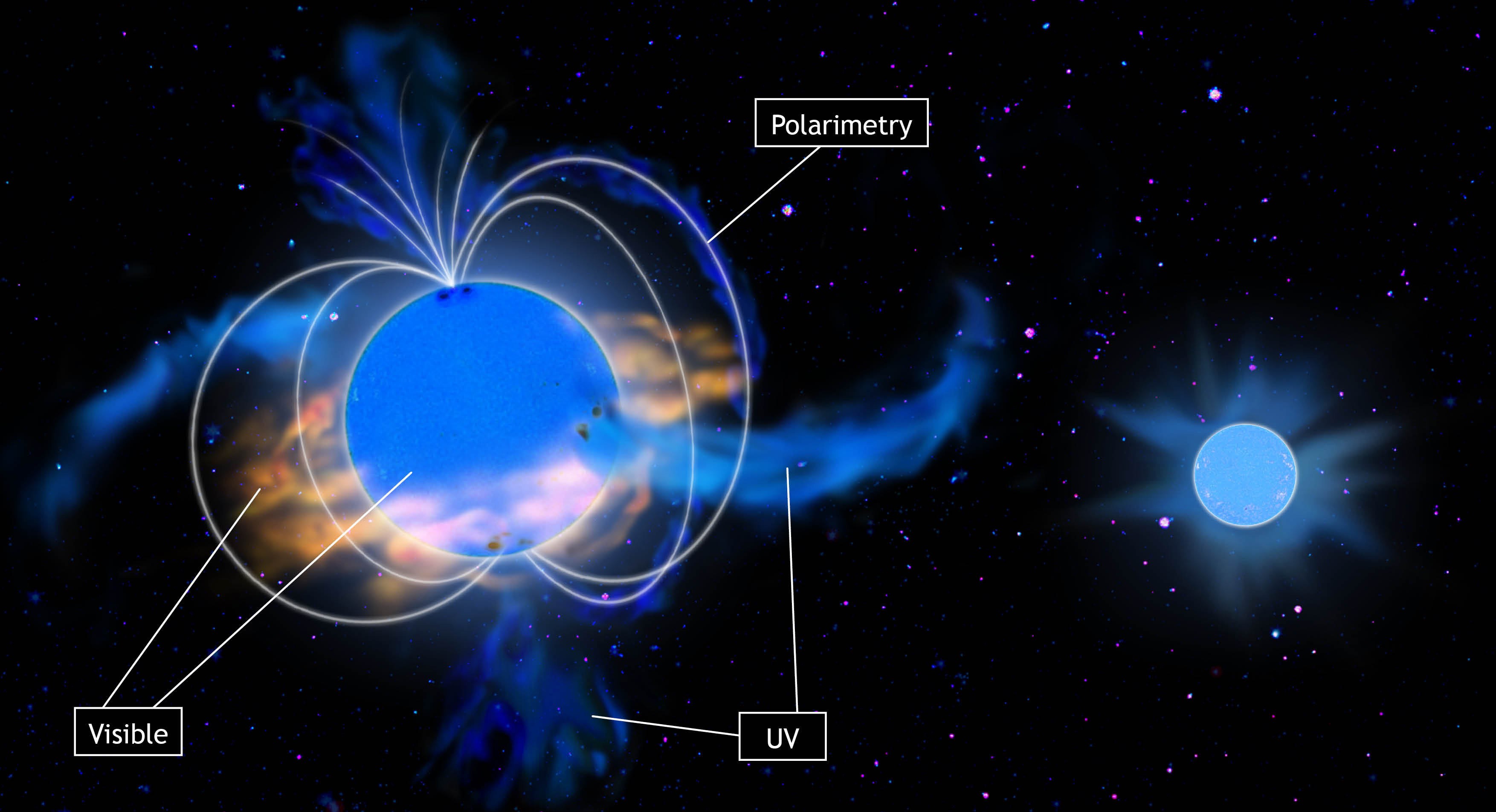}
    \caption{Schematic view of a hot star with its fossil magnetic field lines, channeled polar wind, surface spots, equatorial magnetosphere, corotating interaction regions, and a stellar companion. Copyright: S. Cnudde.}
    \label{fig:hotsketch}
\end{figure*}

\indent Since massive stars emit most of their radiation in the UV domain, and show atomic and molecular lines coming from the photosphere and wind in this wavelength range, these important topics would be best studied with a high-resolution UV spectropolarimeter. UV spectroscopy would indeed allow to characterize the wind and mass-loss properties of hot stars, while the UV polarimetric capabilities would provide, for the first time, a 3D mapping of the magnetized environment. Such maps could be performed on stars with various parameters (rotation, mass,...) and age to understand the impact of magnetic fields along stellar evolution. Moreover, if the UV spectropolarimeter is placed on a sufficiently large telescope, hot stars outside our Galaxy, such as those in the Magellanic Clouds, could be measured as well and allow us to check the impact of metallicity on the results. For those UV spectropolarimetric measurements, high spectral resolution is needed (R$\geq$30000) to allow for sufficient spatial resolution in the 3D maps.

\section{Cool stars and their environments}
\subsection{UV spectroscopy and spectropolarimetry of cool stars}

\indent Cool stars emit only a small fraction of their bolometric luminosity at UV wavelengths and shorter.
A large fraction of the observed UV flux originates from the magnetically-heated upper atmosphere rather than the photosphere. The UV spectra of cool stars therefore represent a unique source of information on their magnetism, upper atmosphere and the way they interact with their environment, complementary with observations at optical wavelengths, in particular.

\subsection{Main sequence stars: chromospheres and impact on planets}

\indent Magnetic fields, their generation through dynamo action and the resulting activity phenomena play a key role in the physics of cool stars and their planetary systems along their whole evolution (Mestel \& Landstreet 2005).\\
\indent With high-resolution spectroscopy and spectropolarimetry at visible and near-IR wavelengths we are now starting to get an overview of the main properties of the photospheric magnetic fields of cool stars from young T~Tauri stars to evolved giants (Donati \& Landstreet 2009). With high-resolution UV spectroscopy/spectropolarimetry it would be possible to extend this picture to the magnetic and thermodynamic structure of the upper atmospheres -- chromospheres and transition regions (TR) -- of cool stars, which is known to be highly structured and dynamic in the solar case thanks to dedicated UV/EUV/X-ray space missions (Del Zanna \& Mason 2018).\\
\indent The UV spectral range indeed contains a number of spectral lines forming at temperatures from $10^4$ to $10^7$~K including chromospheric lines (Mg\textsc{ii}, C\textsc{i}, O\textsc{i}), TR lines (e.g., C\textsc{ii-iv} , N\textsc{iv}, O\textsc{iii-v}, Si\textsc{ii-iv}),
the coronal line Fe\textsc{xii} at 124.2~nm, as well as molecular lines (CO and H$_2$) tracing cool material (Pagano et al. 2004).
From time-series of high-resolution UV spectra, it is possible to reconstruct the structure of the chromosphere (Busa et al. 1999). With polarimetry and a full-UV coverage it would be possible to reconstruct the 3D magnetic and thermodynamic structure of upper stellar atmospheres, connect them with spots and magnetic regions observed at the photospheric level with visible spectropolarimetry, and to constrain models of chromospheric and coronal heating (Testa et al. 2015). Placed on a large aperture telescope, the enhanced sensitivity would allow us to study in details the chromospheres of very low mass stars and brown dwarfs, a key to connect apparently contradictory activity at radio and X-ray wavelengths (Williams et al. 2014); as well as solar-type stars in nearby clusters to build a revised picture of the Sun's activity in time. High-resolution UV spectropolarimetry of cool stars  is also the key to understand how they interact with their planetary systems through stellar winds, flares and associated coronal mass ejections (CMEs), and high energy radiation. These studies will be instrumental in revising the concept of habitable zone beyond the insolation criterion. 

\begin{figure*}[t]
    \centering
    \includegraphics[height=8cm]{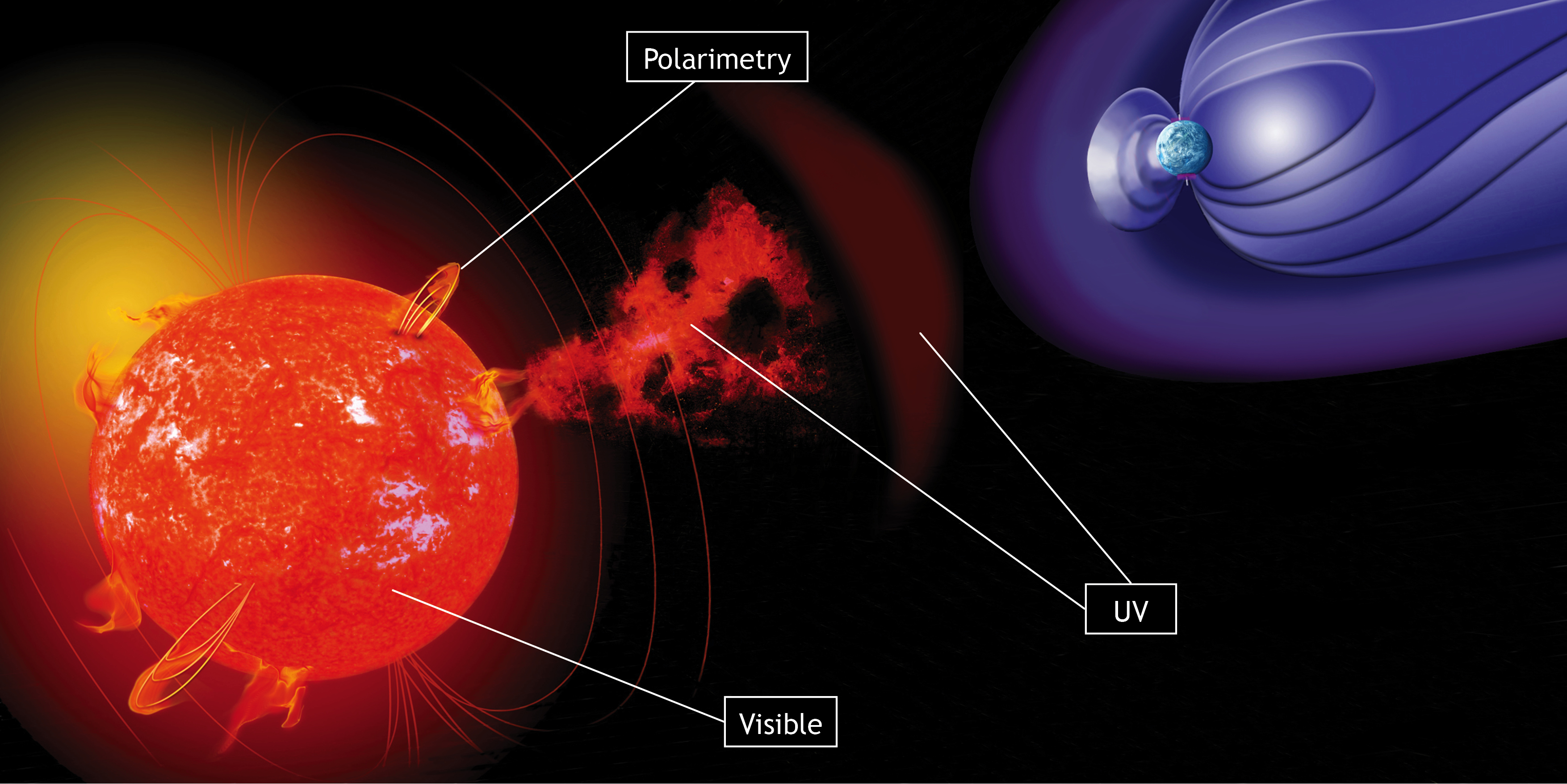}
    \caption{Schematic view of a cool star with its dynamo magnetic field, surface faculae and plages, wind, a coronal mass ejection, and a bow shock between the star and its planet. Copyright: S. Cnudde.}
    \label{fig:coolsketch}
\end{figure*}

\subsection{Pre-main sequence stars: star-disk interaction and accretion-ejection}

In addition to probing stellar magnetism in extremely active regime, high-resolution UV spectropolarimetry of young T~Tauri stars would be a fundamental tool to study the dynamics of accretion shocks and more generally the interaction of the young star with its circumstellar disk. In the magnetospheric accretion model, ionized material is channeled along magnetic field lines onto the stellar surface, and is associated with strong winds and collimated outflows. Gas in the accretion column and in the resulting accretion shock exhibit temperatures in the range $10^4-10^6$~K  emitting a strong blue-ultraviolet continuum along with characteristic emission lines (Hartmann et al. 2016).\\
\indent Despite recent progress in both observations and numerical simulations of accretion flows (Kurosawa \& Romanova 2013), many open questions remain. With time-series of high-resolution UV-visible polarized spectra, it would be possible to study the 3D dynamic structure of accretion shocks and its relation with the surface magnetic field, extending present work in  the visible range (Donati et al. 2012). Such information will be crucial in understanding the relation between stellar parameters, magnetic field properties and the geometry, mass-flux and time-dependence of accretion flows, as well as establishing the physical processes connecting accretion with the wind/outflow process.\\
\indent Observing and understanding high-energy radiation originating from the T~Tauri star is also essential to establish its role in the ionization of the circumstellar disk, and thus in MHD processes such as the magneto-rotational instability, which is generally invoked to explain the observed mass accretion rates. Beyond the accretion/ejection processes, such observations will be of prime importance in devising how this high-energy environment affects the overall properties of the circumstellar disk and the planet formation process.

\subsection{Evolved stars: chromospheres, surface structures and mass-loss}

Cool giant and supergiant stars constitute a late evolutionary stage of low- and high-mass stars respectively. Because of rotational braking during the main sequence and angular momentum conservation during the inflation phase, these stars display long rotation periods and consequently weak dynamo-generated magnetic fields. As in main sequence stars, with high-resolution UV spectropolarimetry it would be possible to study the 3D dynamic structure of the chromosphere of these stars and address the puzzle of chromospheric heating, but in a regime of parameters very different from solar, with very dynamic atmospheres (shocks and pulsations), weak magnetism and active chromospheres often without detectable X-ray coronal counterpart (P{\'e}rez Mart{\'{\i}}nez et al. 2011). It would also be possible to investigate the mechanisms -- pulsations, shocks, turbulence and reduced effective gravity, radiation pressure -- that drive the mass-loss of these stars from the photosphere to the upper atmosphere, as well as the role of magnetic fields (Josselin \& Plez 2007); while mass-loss would be simultaneously measured through far-UV lines (Dupree et al. 2005).\\
\indent 
Finally, the recent discovery of linear polarisation in the spectral lines of cool supergiants -- particularly strong in the blue part of the visible spectrum -- stimulates the development of novel techniques to study the atmospheric dynamics of these stars (Lopez et al. 2018). Extending such methods to a simultaneous UV-visible coverage appears particularly promising to constrain the link between the large-scale supersonic convection observed on these stars and mass-loss.

\section{Conclusions}

No space mission equipped with a high-resolution UV spectropolarimeter covering a wide wavelength domain has ever flown. Such an instrument, in particular on a large aperture space telescope, would open a new door in stellar physics by allowing to investigate the origin, evolution, and impact of magnetic fields and magnetospheric structures in both hot and cool stars. This would in particular allow us to take a leap forward in the comprehension of the structure, environment, activity, and evolution of all types of stars, with important consequences on many other domains of astrophysics.

\pagebreak
\textbf{References}

- Alecian, E., Neiner, C., Wade, G. A. et al. 2015, IAUS 307, 330

- Blazère, A., Petit, P., Lignières, F. et al., 2016, A\&A 586, A97

- Bus{\`a} I., Pagano I., Rodon{\`o} M., Neff J.~E. \& Lanzafame A.~C., 1999, A\&A 350, 571 
  
- Commerçon, B., Hennebelle, P. \& Henning, T., 2011, ApJ 742, L9

- Del Zanna G. \& Mason H.~E., 2018, LRSP 15, 5

- Donati J.-F. \& Landstreet J.~D., 2009, ARA\&A 47, 333 

- Donati J.-F., Gregory, S. G., Alencar, S. H. P. et al., 2012, MNRAS 425, 2948 

- Dupree A.~K., Lobel A., Young P.~R. et al., 2005, ApJ 622, 629 

- Featherstone, N. A., Browning, M. K., Brun, A. S. \& Toomre, J., 2009, ApJ 705, 1000

- Fossati, L., Schneider, F. R. N., Castro, N. et al., 2016, A\&A 592, A84

- Grunhut J. \& Neiner C., 2015, IAUS 305, 53

- Guilet, J., Foglizzo, T. \& Fromang, S., 2010, SF2A 2010, 173

- Hartmann L., Herczeg G. \& Calvet N., 2016, ARA\&A 54, 135 

- Josselin E. \& Plez B., 2007, A\&A 469, 671   
  
- Kurosawa R. \& Romanova M.~M., 2013, MNRAS 431, 2673 

- L{\'o}pez Ariste A., Mathias, P., Tessore, B. et al., 2018, A\&A 620, A199 

- Mestel L. \& Landstreet J.~D., 2005, LNP 664, 183 
 
- Meynet, G., Eggenberger, P. \& Maeder, A., 2011, A\&A 525, L11

- Neiner, C., Mathis, S., Alecian, E. et al., 2015, IAUS 305, 61

- Osten R.~A., Brown, A., Ayres, T, R. et al., 2004, ApJS 153, 317 

- Pagano I., Linsky J.~L., Valenti J. \& Duncan D.~K., 2004, A\&A 415, 331 

- P{\'e}rez Mart{\'{\i}}nez M.~I., Schr{\"o}der K.-P. \& Cuntz M., 2011, MNRAS 414, 418 

- Petit V., Keszthelyi, Z., MacInnis, R. et al. 2017, MNRAS 466, 1052 
  
- Testa P., Saar S.~H. \& Drake J.~J., 2015, RSPTA 373, 20140259 

- Williams P.~K.~G., Cook B.~A. \& Berger E., 2014, ApJ 785, 9 

\end{document}